# DETERMINANTS OF MIGRATION: A SIMPLE LINEAR REGRESSION ANALYSIS IN INDIAN CONTEXT

**Soumik Ghosh\*\*, Arpan Chakraborty\***

\*\* M.Phil at IGIDR

\*PhD Research Scholar at IIT Kharagpur

EMAIL ID:  soumik@igidr.ac.in ; arpan.ms97@kgpian.iitkgp.ac.in

## Abstract:

Environmental degradation, global pandemic and severing natural resource related problems cater to increase demand resulting from migration is nightmare for all of us. Huge flocks of people are rushing towards to earn, to live and to lead a better life. This they do for their own development often ignoring the environmental cost. With existing model, this paper looks at out migration (interstate) within India focusing on the various proximate and fundamental causes relating to migration. The author deploys OLS to see those fundamental causes. Obviously, these are not exhaustive cause, but definitely plays a role in migration decision of individual. Finally, this paper advocates for some policy prescription to cope with this problem.

**Keywords:**  Migration, proximate cause of migration

# INTRODUCTION:

The very thought of migration is an age-old concept practiced by humanity to prevent adversity. From the ages of civilization to present day it's still persistent with the cause changing accordingly. Ranging from preventing harsh climate, enemies to economic slowdown and policy failure in native place stimulates migration. However, with population explosion the new trend of mass migration raises questions on sustainable development and serious Environmental concerns. It also exacerbates regional inequality within a country and also may lead to weaker bilateral ties. With 258 million total migration (68 million solely due to war) across country in global scenario the UN Summit meet of 2017 addressed the migration policy and stressed for investigation of its root cause.

With India not lagging behind, 30% (Census 2015) of its population engage themselves in this issue mostly in terms of job search and better education facility. Out of that 98% itself is intra country migration indicating a greater association with regional inequality. Without lake of urban facility in towns and failure of spreading urban benefits across the whole terrain is a menace to the developing country like ours. The words of economist John W Mellar "Improper pattern of urbanization leading to chaotic migration in developing country" clearly indicates a bureaucratic plan failure in this regard.

# LITERATURE REVIEW:

Numerous neoclassical models have been built to study the migration both in and out behavior in an economy. First of all, Arthur Lewis (1955) proposed the dual sector model where growth of an underdeveloped country hinges on migratory worker's decision to move from subsistence to heavy capital industry assuming equal wage. Harris & Todaro (1970) jointly criticized this decision saying the decision depends on expected wage difference between origin and native country. Much a holistic model of migration was proposed by Lee (1966) taking into account the push factor at origin and pull factor at destination. These act intermittently as disincentive and incentive for migration decision to be based upon. The push factor is low wages, unemployment level political stabilization compared to its contradictory level at destination which may bring about migration. These were mostly broad issues of migration without looking into the internal migration issues. The first of this concept was raised in Zelinsky et al. (1971) he discussed about a model which even focuses on the aspect of urbanization development within a country. He encompassed the rate of migration with the onslaught of urbanization. Subsequently the widening of regional inequality. The outstanding feature of his model is the way he divided each into 5 stages of development and showed the trend of migration. The main conclusion from his model was intraregional migration is more likely stimulated by economic cause. Perhaps the most comprehensive theory of migration was investment in human capital theory given by Saajstad (1962) where he says migration is a decision coming out from cost – benefit analysis of moving from one place to other. From the theoretical work the author has encapsulated the ideas and have tried to make an empirical estimation of short distance or internal migration of India across states while keeping into mind about the economic factors in mind. Also, keeping the evidences in mind as well as keeping in mind about the secondary data sources, the author only used a few important determinant variables. The first priority of the paper is to provide greater economic benefit and to empirically analyze the root causes related to migration. In support with these econometric work by Greenwood (1975) and Kria et al. (1996) which were successful and these papers are related to this paper. But obviously, this work is unique in the sense of estimation and data related micro foundations of migration.

However as much of the empirical research goes the scenario is not that much clear in different parts of the globe in modern times. Bukeniya et al. (2003) examined the effect of wage differential and regional job growth in West Virginia. With the correlation between employment opportunities and wage differential a clear picture is quite difficult to get often giving rise to endogeneity. They showed that though better economic outcome provides the incentive but decision is based mainly on geographical location often proving lesser dependence on wage or employment opportunities. Nanfoso and Akono (2003) while carrying out a similar experiment in urban Cameroon observed that in spite of high wage rate and better opportunities speculative migration to urban Cameroon remains unexplained. Barro et al. (1991) proved that most empirical work on net money wage differential would not affect much of migration decision through a neoclassical production function. The result was 10 % increase in per capita income would give rise to only 0.26 % of migration decision. Venti and Wise (1984) showed money wage differential to provide a very weak incentive for migration. Mostly, this arises due to resettlement cost. Thus, the new genre of migration decision after the development of neoclassical model no longer focuses on these two variables. In a study by Mukherji - DPFW (1993) its showed that interstate migration is positively correlated just for survival issues like development and employment and not on education purposes. While mostly the migration tends to be circular, the level of environmental degradation is scary due to overburden on urban natural resources, thus necessitating the need to find the root of the evil Deshingkar and Start (2004). Similarly, Bhagat et al. (2010) has showed increased population mobility from post reform period.

Now a days migration decision is based on family endeavors. Migrants migrate to places well connected with familiar ties. Works of Lucas (1997) and Winters (2001) are clear evidence of it. Thus, the author runs a regression taking into account only the simple causes of migration which can be found logical from the literature and we have data for that. This paper can be also be used for future work related to endogenous macro theory of migration.

**THE DATA:**

Obtaining the data of migration is always a challenge. Most of the data source are insufficient in even stating the externality effect. This can have large stimulus effect on regression results. The transient nature of data is also another important thing to look whenever we talk about migration related dataset. However, we have managed to procure the data set from IHDS (India Human Development Survey) 2011– 13 rounds. This was the second round of data collection with the first round being in 2005. The data source contained number of migrations in the last 5 year. We specifically used the DSC001 folder data from IHDS website which comes directly in .dta file. We import the data to Stata software.

MG4 variable contains data regarding the destination country of the migrant. MG4 equaled 1 and 2 for same state or another state migration, while 3 was for abroad which was also considered for our analysis. We do this in order to check the interstate migration and the proximate causes which we often see from usual labor economics related literature. MG5 was the variable containing relevant information regarding the type of city the migrant is heading. It was also a binary variable for urban and rural type. If rural then 0 or 1 otherwise. We consider MG4 as my dependent variable and rest MG5, MG6, MG7, MG8 as independent variables.

The model adopted here is a simple OLS model as written below,

$$Y = \delta + \alpha A + \beta B + \theta C + \gamma D + \epsilon$$

Where Y = number of interstate migrants (within India moving from one state to another)

A = MG5 (rural or urban area)

B = MG6 (gone alone or with family)

C = MG7 (gone for how many months)

D = MG8 (gone for how many years ago), **with the usual intercept term δ,**

In order to make our model simple and realistic, we use the OLS model. In the result section we will see the probable problems which might happen because of OLS estimation. Please refer to our model in table 1.1 and please note the notations used as A, B, C, D their descriptions follow from the above regression model.

**Table1.1: Regression Coefficients of OLS**

| <u>VARIABLE</u> | <u>COEFFICIENT with p value</u> |
|---|---|
| A | 0.199123(0.000) |
| B | -0.05254 (0.000) |
| C | 0.004623(0.000) |
| D | -0.0031806(0.69) (Not significant) |
| constant | 1.494987(0.000) |

Result by Author. P-values are in brackets.

R – Square = 0.0491

The coefficients of the regressors are very small in magnitude however they all are statistically significant with an extreme low value of R Square except 'D'. The significance part shows the extent of effect where the data of migrant and its causes are obtained where there is a relation between the two. Thus, we can conclude that these regressors do have a role to play. Most of them are significant at 99 % level of significance except for how many years ago the migrant has gone, which eventually is not significant. But it does not conclude that this is unimportant variable. So, the author did not exclude it from the model. May be, we need to find an instrumental data for specially MG5 and MG4, and re run the regression. But unfortunately, we do not have a readymade instrumental variable data. The author will surely work on this in future work.

However, with statistical complicacy, it shows that there is a bias from simultaneous equation creating endogeneity problem (the variables may be both simultaneously determined, specifically for MG4 and MG5, i.e., the rural urban migration and interstate migration in India). It can even be sensed at very low R squared value. But our data satisfy consistency property and shows the no presence of heteroscedasticity (Breusch Pagan test). Also, in Stata, the author ran Cameroon Trivedi's autocorrelation (known as generalized white test in the literature) test hence our data is not autocorrelated. But our hypothesis can be as follows, the best treatment to get a clearer picture would be to either use IV or fixed effect estimation models. From here we can also observe the effect of geographical mobility (probably) playing a part through the unobservable that crept in our model. The results of the Breusch Pagan test and Cameroon Trivedi test are given below:

**Table1.2: Cameroon Trivedi Test**

```
White's test for Ho: homoskedasticity
         against Ha: unrestricted heteroskedasticity

         chi2(13)      =      367.06
         Prob > chi2   =      0.0000

Cameron & Trivedi's decomposition of IM-test
```

| Source | chi2 | df | p |
|---|---|---|---|
| Heteroskedasticity | 367.06 | 13 | 0.0000 |
| Skewness | 147.38 | 4 | 0.0000 |
| Kurtosis | 420.98 | 1 | 0.0000 |
| Total | 935.43 | 18 | 0.0000 |

### Table1.3: Breusch Pagan Test

```
Breusch-Pagan / Cook-Weisberg test for heteroskedasticity
         Ho: Constant variance
         Variables: fitted values of MG4

         chi2(1)      =     32.52
         Prob > chi2  =    0.0000
```

### **LIMITATION OF THE STUDY:**

Given the statistical results of this paper, we are not in a position to run instrumental variable estimation for robust estimation which can be sensed from our R square of the model. In our future work we can collect primary data and run the same regression model and then we can justify our analysis.

### **CONCLUSION & POLICY PRESCRIPTION:**

With our model, we are able to justify from the secondary data the determinates of inter-state migration using a simple linear regression. Besides the paper shows that there is a role to play on the part of bureaucrats to promote ethnic integrity in India, because these omitted variables in regression can have huge impact over our low R-square. Also, with the limited model capacity, this analysis should be only accomplished only when all the probable destination of the migrant are equal in both economic and humanitarian terms. Then only the problem of environmental degradation arising from migration can be mitigated making metropolitan cities and towns a healthy place to live in. However, this study can surely indicate the bureaucrats to focus on the micro motives of migration which is the success of this study.

# **REFERENCES:**

- Barro, R. J., and X. Sala-i-Martin. Regional Growth and Migration: A Japan-U.S. Comparison. Working Paper. *National Bureau of Economic Research* (1991).

- Bhagat, R.B (2010): 'Internal migration in India: are the underprivileged class migrating more?' *Asia-Pacific Population Journal*, Vol 25, No1, pp 27-45.

- Bukeniya, James et al. "The effect of wage differential and job growth on migration: A case on West Virginia" (2003)

- Deshingkar, P. and D. Start 2003 Seasonal Migration for Livelihoods, Coping, Accumulation and Exclusion, Working Paper No. 220, *Overseas Development Institute*, London

- Greenwood, M. J. (1975). Research on internal migration in the United States: a survey. *Journal of Economic Literature*, 13, pp.397-433

- Harris, J. R. & Todaro, M. P. (1970). Migration, unemployment and development: a two-sector analysis. *American Economic Review*, 60, pp.126-142.

- Kriaa, M., & Plassard, J. M. (1996). La mobilité géographique des diplômés de l'enseignement supérieur français : processus de double de gains. *Recherches économiques de Louvain*, *62* (1), pp.95-122.

- Lucas, R. (1997). Internal migration in developing countries. In Rosenzweig, M. R., & Stark, O. (eds), *Handbook of Population and Family Economics*, Elsevier Science BV.

- Mukherji, S. (1991); "The Nature of Migration and Urbanization in India: A Search for Alternative Planning Strategies," *Dynamics of Population and Family Welfare*, Mumbai, PP. 203-245

- Nanfoso, Roger ; "Migration and wage differential in urban Cameroon" *Research in Applied Economics* 2009 vol. 1

- Sjaastad, L. A. 'The Costs and Returns of Human Migration.' *Journal of Political Economy* 70, vol. no. 5(1962): 80-93

- Venti, S. J., and D. A. Wise. 'Moving and Housing Expenditure: Transaction Costs and Disequilibrium.' *Journal of Public Economics* 23(1984): 207-243.

- Winters, P., de Janvry, A., & Sadoulet, E. (2001). Family and community networks in Mexico-US migration. *Journal of Human Resources*, *36* (1), pp.159185

- Zelinsky, Wilbur (April 1971). "The Hypothesis of the Mobility Transition". *Geographical Review*. **61** (2): 219–249. doi:10.2307/213996. JSTOR 213996.